**Optical properties of $Cu_2ZnSn(S_xSe_{1-x})_4$ by first-principles calculations**

Sergiy Zamulko[1], Kristian Berland[1], and Clas Persson[1,2]
[1]Centre for Materials Science and Nanotechnology, University of Oslo, P. O. Box 1048 Blindern, NO-0316 Oslo, Norway
[2]Department of Materials Science and Engineering, KTH Royal Institute of Technology, Brinellvägen 23, SE-100 44 Stockholm, Sweden

Email address of the corresponding author: sergii.zamulko@smn.uio.no

**ABSTRACT**

Structural, electronic and optical properties of $Cu_2ZnSn(S_xSe_{1-x})_4$ semiconductors are studied theoretically for different concentration of S and Se anions. The optical properties are calculated at three levels of theory, in the generalized gradient approximation (GGA), meta-GGA, and with a hybrid functional. The GGA and meta-GGA calculations are corrected with an on-site Coulomb $U_d$ term. Lattice constants, dielectric constants, and band-gaps are found to vary almost linearly with the concentration of S. We also show that a dense sampling of the Brillouin zone is required to accurately account for the shape of the dielectric function, which is hard to attain with hybrid functionals. This issue is resolved with a recently developed $\mathbf{k}\cdot\tilde{\mathbf{p}}$ based interpolation scheme, which allows us to compare results of the hybrid functional calculations on an equal footing with the GGA and meta-GGA results. We find that the hybrid functionals provide the overall best agreement with the experimental dielectric function.

**INTRODUCTION**

The kesterite-type phases of $Cu_2ZnSnS_4$ (CZTS), $Cu_2ZnSnSe_4$ (CZTSe) and mixtures of these (here called CZTSSe) have attracted attention from the scientific and industrial community as potential earth-abundant and inexpensive solar cell materials [1]. Much progress has been made on understanding how the synthesis affect their properties such as material stability and conductivity [2-6], and the synthesis affects device performance [7]. The optical properties of CZTS [8,9] and CZTSe [9,10] compounds have been studying extensively [6,9]. There has also been a number of computational [11-17] and experimental [2-7,18] studies addressing structural, electronic, and optical properties of their mixtures, i.e. CZTSSe. The highest efficiency attained so far for CZTSSe [19] is 12.6%, which was achieved by partially substituting S anions with Se [5-7,11,18,20] forming a band-gap grading through the depth of the film with optical gaps ranging from 1.1 to 1.5 eV.

In assessing strategies to enhance photovoltaic efficiency, first-principles calculations of optical properties can be used to optimize the atomic substitution and in the analysis of optical measurements. For instance, Camps *et al.* [14] calculated using density functional theory (DFT) the absorption coefficient of kesterite and stannite $Cu_2ZnSn(Se_xS_{1-x})_4$ in the local density approximation with 64 atom supercells. Recently, Li *et al.* [21] calculated the dielectric function of the kesterite phase using the modified meta-GGA of Becke-Johnson (mBJ) [22,23] functional for an 8 atom supercells and a $20 \times 20 \times 20$ **k**-mesh. There has also been a number of hybrid functional calculations [15,24-27], in the Heyd-Scuseria-Ernzerhof variant of 2006 (HSE06) [28]. For a number of inorganic compounds, hybrid functionals have been shown to exhibit higher accuracy [27,29,30] than computationally inexpensive calculations based on semi-local exchange-correlation functional. For $Cu_2ZnSnS_4$ and $Cu_2ZnSnSe_4$ compounds, using HSE06 improves the account of band-gaps and magnitude of dielectric functions. A. Crovetto *et al.* demonstrated for $Cu_2SnS_3$ that a **k**-point sampling of 30×30×30 (6992 **k**-points) was required to accurately resolve the shape of the imaginary dielectric functions close to the band edge [31].

However, because of the high computational costs, all the hybrid functional calculations have been limited to a coarse sampling of the Brillouin zone; for instance, Sarmadian *et al*. [15] used a $4 \times 4 \times 4$ mesh. Thus, a numerically converged comparison of the accuracy of different exchange-correlation functionals in DFT is missing. For the same reason, there has been no comparison between hybrid functional calculations and experimental data for characteristic features of the dielectric function for different S-Se ratios.

In this paper, we overcome the **k**-point sampling issues for hybrid functionals by employing a recently developed, by study by employing a recently developed $\mathbf{k} \cdot \mathbf{p}$ -inspired interpolation scheme (the $\mathbf{k} \cdot \tilde{\mathbf{p}}$ method) [32,33], which allows us to compare, on an equal footing, how hybrid and semi-local exchange-correlation functionals in density functional theory (DFT) affects the dielectric function of CZTSSe.

In particular, we compare the dielectric function of HSE06 with the results with those of the generalized gradient correction (GGA) of Perdew-Burke-Ernzerhof (PBE) [34] and the mBJ, both corrected with an *ab initio*-derived on-site Coulomb $U_d$ term. The dielectric constants obtained with HSE06 and mBJ+$U_d$ functionals agree better with experiment than PBE+$U_d$ even when including scissor corrections. Moreover, while HSE06 can provide the band-gap value within 0.3 eV to experimental without using any corrections, PBE+$U_d$ can deviate by as much as 1 eV eV. mBJ+$U_d$ gives a band-gap within 0.36 eV of the experimental. However, even when comparing the dielectric function fixed to the same band-gap, the real and imaginary part of the HSE06 dielectric function has the best agreement with experiment for the dielectric function, as characterized by the overall magnitude, the peak position in the imaginary part and the position of the kinks in the real part of dielectric function. For instance, mBJ+$U_d$ with scissor correction could result in a peak position 0.71 eV above the experimental, whereas HSE06 is always within 0.32 eV. Being able to accurately predict such characteristic features of the dielectric function is essential in the analysis of solar cell materials, such in the extraction of accurate optical band-gaps from Tauc absorption plot, as well as for obtaining accurate dielectric constants, which is widely used for solar cell modeling.

## THEORY

### $\mathbf{k} \cdot \tilde{\mathbf{p}}$ method

Dense sampling of the Brillouin zone is required in computing a range of material properties, such as in electronic transport. Several wave-function based interpolation schemes have been developed to interpolate the Brillouin zone, including Wannier [35-37] and Shirley interpolation [38,39]. Here, we employ the recently developed $\mathbf{k} \cdot \tilde{\mathbf{p}}$ method [32,33], which is a simple and accurate interpolation method. It builds on the standard extrapolative $\mathbf{k} \cdot \mathbf{p}$ method [32,33,35], but introduces a correction term that resolve band crossings and enhance accuracy, making it suitable for interpolating between **k**-points. We recently updated the method[33] to rely on the velocity-matrix element rather than momentum-matrix elements [40], enhancing the accuracy for  non-local one-electron potentials, such as arising from the Fock term in hybrid functionals.

### Computational details

First-principles calculations are performed with the projector augmented wave (PAW) method as implemented in the Vienna *ab initio* simulation package (VASP) [41]. Electronic band dispersion is obtained using PBE+$U_d$, mBJ+$U_d$, and HSE06. $U_d$ is set to 4.0 eV for Cu-3*d* and 7.5 eV for Zn-3*d* orbitals [42]. Lattice and atomic coordinates are relaxed using the PBEsol functional [34]. The mixing enthalpy of $Cu_2ZnSn(S_xSe_{1-x})_4$ is calculated with a 64-atom supercell for $x$ = 0, 0.125, 0.25, …, and 1. The plane wave energy cutoff is set to 450 eV throughout this study.

Optical properties and band structures are calculated with an eight-atom supercell with $x$ = 0, 0.25, 0.5, 0.75, and 1. The total energy is converged to $10^{-8}$ eV and $10^{-6}$ eV for respectively 8- and 64- atom unit cells with forces relaxed to $5\cdot10^{-3}$ eV/Å. In the structural relaxation, a 4 × 4 × 4 (10 × 10 × 10) Γ-centred **k**-mesh is used for the 64 (8) atom cells with a quasi-random distribution of S/Se anions. The dielectric function is calculated with a 24 × 24 × 24 **k**-mesh for PBE+$U_d$ and mBJ+$U_d$ and 64 bands, which converge $\varepsilon_\infty$ within 0.1. For HSE06, we employ an 8 × 8 × 8 **k**-mesh (we denoted this as $\mathbf{k} = 8^3$) interpolated to a 24 × 24 × 24 **k**-mesh (we denoted this as $\mathbf{k} = 24^3$) using the $\mathbf{k}\cdot\tilde{\mathbf{p}}$ interpolation scheme [32,33] as exemplified for PBE+$U_d$ in Fig. 2 $\varepsilon_2$ is computed with an in-house implementation of the linear-tetrahedron method [43].

## RESULTS AND DISCUSSION

### Structural properties and mixing enthalpy.

The calculated lattice constants (both $a$ and $c$) of kesterite $Cu_2ZnSn(S_xSe_{1-x})_4$ underestimate the experimental [44,45] by up to 0.05 Å (Table 1), as do earlier calculations employing PBEsol [17]. In agreement with Vegard's law, both experimental and calculated lattice increase almost linearly with Se-content from calculated values of $a$ = 5.374 and $c$ = 10.751 Å for $Cu_2ZnSnS_4$ to 5.661 and 11.318 Å for $Cu_2ZnSnSe_4$. The expansion can be related to that the larger atomic covalent radius of Se (1.20 Å) than that of S (1.05 Å).

The mixing enthalpy of the $Cu_2ZnSn(S_xSe_{1-x})_4$ solid solution is given by

$$\Delta H(x) = E(x) - [(1-x)E(0) + xE(1)] \quad (1)$$

**Table 1.** Calculated lattice parameters ($a$ and $c$), and positions of anions of $Cu_2ZnSn(S_xSe_{x-1})_4$ for the alloy composition $x$ = 0, 0.25, 0.5, 0.75 and 1, obtained from modeling 8 atom primitive cells

| Compound | $a$ [Å] PBEsol/ other calc./ exp | $c$ [Å] PBEsol/ other calc./ exp | positions of anions, $x$ $y$ $z$ |
|---|---|---|---|
| $Cu_2ZnSnS_4$ | 5.374/5.365 [17]/ 5.428 [27] | 10.751/10.739 [17]/10.867 [27] | S(1): 0.111 0.139 0.240<br>S(2): 0.648 0.620 0.240<br>S(3): 0.860 0.351 0.760<br>S(4): 0.379 0.888 0.760 |
| $Cu_2ZnSnS_3Se_1$ | 5.440/5.417 [17]/ | 10.89710.843 [17]/ | S(1): 0.113 0.136 0.239<br>S(2): 0.645 0.619 0.239<br>S(3): 0.860 0.349 0.758<br>Se(1): 0.380 0.892 0.762 |
| $Cu_2ZnSnS_2Se_2$ | 5.513/5.470 [17]/ | 11.032/10.947 [17]/ | S(1):0.110 0.136 0.241<br>S(2): 0.858 0.352 0.758<br>Se(1): 0.652 0.618 0.236<br>Se(2): 0.376 0.889 0.762 |
| $Cu_2ZnSnS_1Se_3$ | 5.594/5.521 [17]/ | 11.180/11.052 [17]/ | S(1): 0.113 0.138 0.240<br>Se(1): 0.649 0.622 0.236<br>Se(2): 0.855 0.346 0.761<br>Se(3): 0.379 0.893 0.760 |
| $Cu_2ZnSnSe_4$ | 5.661/5.574 [17]/ 5.68 [27] | 11.318/11.183 [17]/ 11.360 [27] | Se(1): 0.110 0.141 0.238<br>Se(2): 0.651 0.620 0.238<br>Se(3): 0.858 0.348 0.761<br>Se(4): 0.379 0.889 0.761 |

where $E(1)$ and $E(0)$ are the total energies of $Cu_2ZnSnS_4$ and $Cu_2ZnSnSe_4$ respectively, and $E(x)$ is the total energy of $Cu_2ZnSn(S_xSe_{1-x})_4$. $\Delta H(x)$ can be fitted to a quadratic function of composition $x$ [16], as follows

$$\Delta H(x) = (1-x)\Delta H(0) + x\Delta H(1) + \Omega x(1-x) \quad (2)$$

where $\Omega$ is the alloy mixing parameter. We find a value of $\Omega = 24.4$ meV, whereas as Zhao *et al.* [17], reports a value of 28.5 meV and Chen *et al.* [16] reports 26.0 meV. Figure 1 compares our mixing result with theirs [16,17]. The calculations of Chen. *et al* [16] were based on PW91 [46] in VASP, while Zhao *et al.* [17] used PBEsol, as in our study, but with the Cambridge Serial Total Energy Package [47].

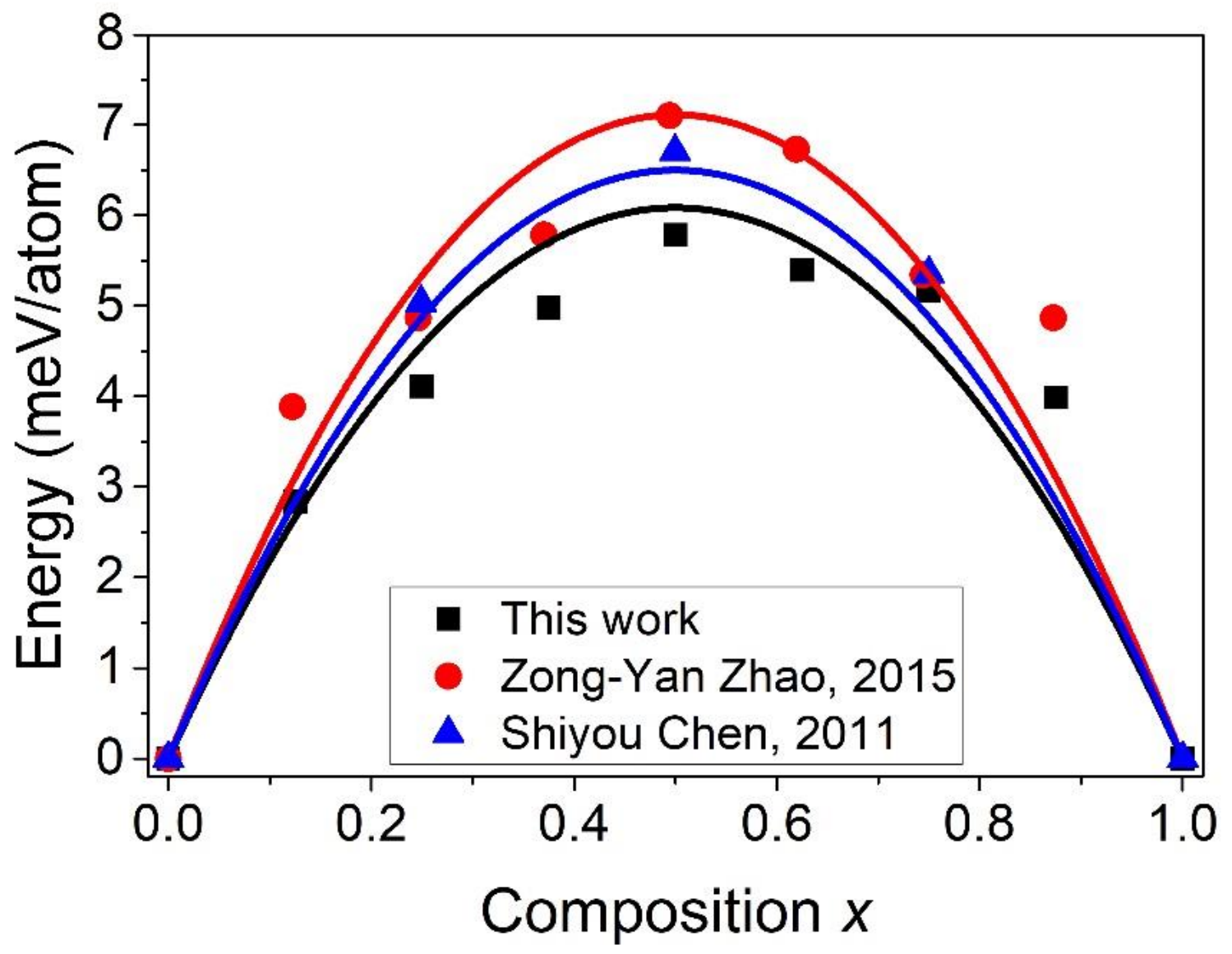

FIG.1. (Color online) Calculated mixing enthalpy of $Cu_2ZnSn(S_xSe_{1-x})_4$ and corresponding fitting curves compared with results in literature (Zhao *et al.* [17] and Chen *et al.* [16]).

**$\mathbf{k} \cdot \tilde{\mathbf{p}}$ method**

Figure 2 illustrates that the $\mathbf{k} \cdot \tilde{\mathbf{p}}$ method can accurately reproduce the density of states and imaginary dielectric function $\varepsilon_2$ of $Cu_2ZnSnS_4$. It compares PBE+$U_d$ results with a $24 \times 24 \times 24$ Γ-centered Monkhorst-Pack sampling (gray background) with results based on an $8 \times 8 \times 8$ mesh (solid curve), results of the $\mathbf{k} \cdot \tilde{\mathbf{p}}$ method interpolated for an $8 \times 8 \times 8$ mesh to a $24 \times 24 \times 24$ mesh (dashed curve). As a test case, we also show in Fig. S1 in the supplementary information (SI) that the Kohn-Sham solution computed with a $4 \times 4 \times 4$ mesh using HSE06 can rather accurately reproduce the dielectric function calculated generated with an $8 \times 8 \times 8$ mesh.

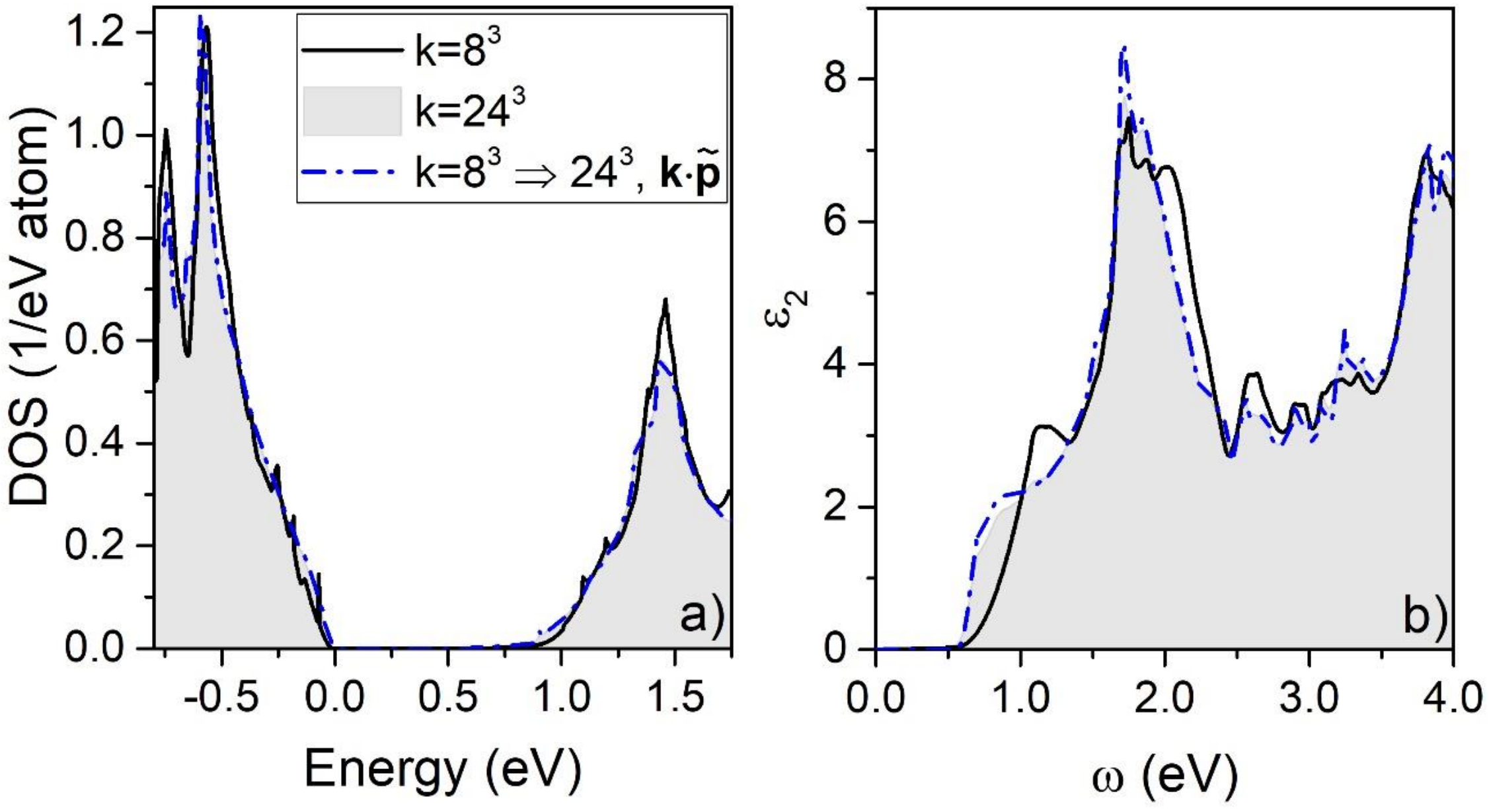

FIG.2. (Color online) The density of states (a) and imaginary part of the dielectric function (b) of $Cu_2ZnSnS_4$ calculated with PBE+$U_d$ with an 8 × 8 × 8 ($\mathbf{k}$ = $8^3$) and 24 × 24 × 24 ($\mathbf{k}$ = $24^3$) Γ-centered **k**-point sampling. The $\mathbf{k} \cdot \tilde{\mathbf{p}}$ method is used to extend **k**-grid from $\mathbf{k} = 8^3$ to $\mathbf{k} = 24^3$.

Panel b) in Fig.2 shows that dense sampling is needed for accurately computing $\varepsilon_2$ close to the band edge. $Cu_2ZnSnS_4$ has a direct band-gap, but with an 8 × 8 × 8 mesh, the dielectric function has a parabolic shape, which is characteristic of weak of dielectric response for photons with energies just above the band-gap. Such a result often is contradicted with corresponded experimental observations and can be explained as an artifact of the effectively linear band approximation in the linear-tetrahedron method at low **k**-sampling.

**Optical properties**

The calculated band-gap of $Cu_2ZnSn(S_xSe_{1-x})_4$ with $x$ = 0, 0.25, 0.5, 0.75, and 1 are presented in Fig. 3 (detailed in Table S1). The band-gaps calculated with HSE06 and mBJ+$U_d$ functionals differ by no more than 0.1 eV and both underestimate the experimental gap by about 0.3 eV [21,27]. For $Cu_2ZnSnSe_4$, the PBE+$U_d$ gap is merely 0.1 eV compared to an experimental gap of 1.0 eV [24]. The band-gaps of all three methods increase with S substitution, in agreement with earlier studies [16,17]. The trend can be fitted to a parabolic form $E_g(x) = (1-x)E_g(0) + xE_g(1) - b \cdot x(1-x)$, where $E_g$ is the band-gap and $b$ is the bowing parameter. Fig. 3. shows that the fits are almost linear, with $b$ values of 0.06 eV, 0.09 eV and 0.003 eV for PBE+$U_d$, mBJ+$U_d$, and HSE06. Such low values are also reported in Refs. [11,17]. The small bowing parameter also reflects the good miscibility of the solid solution [17].

Figure 4 shows the band structures along the symmetry directions Γ-X (100) and Γ-Z (001). $Cu_2ZnSn(S_xSe_{1-x})_4$ has a direct gap at the Γ-point. In the figure, the conduction bands (CBs) of PBE+$U_d$ (full curves) have been energetically shifted by ($E_g^{\mathrm{HSE06}} - E_g^{\mathrm{PBE+U}_d}$) to match the gap of HSE06. Except for the underestimation of the gap, the band structure of PBE+$U_d$ are similar to that of HSE06 (dots).

Since the gaps $E_g$, of HSE06, PBE+$U_d$, and mBJ+$U_d$ are lower than the experimental gaps [21], we compute the dielectric function with a constant shift of the band-gap, widening it to the linear fit to the experimental data of Li *et al.* [21] (Fig. 3 solid line $f(x) = 1.0486 + 0.51895x$). Values of the constant shifts $\Delta_g$ are provided in Table S2.

In Fig. 5, we compare experimental and calculated dielectric constants of $Cu_2ZnSn(S_xSe_{1-x})_4$, with $x$ = 0, 0.25, 0.5, 0.75 and 1 for mBJ+$U_d$ and HSE06 based on coarse and dense Brillouin zone sampling, with (b) and without (a) a $\Delta_g$ constant shifts widening the band-gaps. Introducing a constant reduces the dielectric constants. Whereas standard PBE+$U_d$ overestimate the dielectric constants by up to 74%, this is reduced to 18% with constant shifts. The static- and high-frequency constants differ by about $\varepsilon_0 - \varepsilon_\infty \approx 3$ (Table S1 of SI), indicating a moderate iconicity, similar to that of ZnSe [49]. In line with the empirical Moss-Ravindra relations [50], the dielectric functions decrease with increasing band-gap and hence with S substitution. However, the dielectric constants of the various compositions differ by no more than 2 for all but the PBE+$U_d$ results without constant shifts.

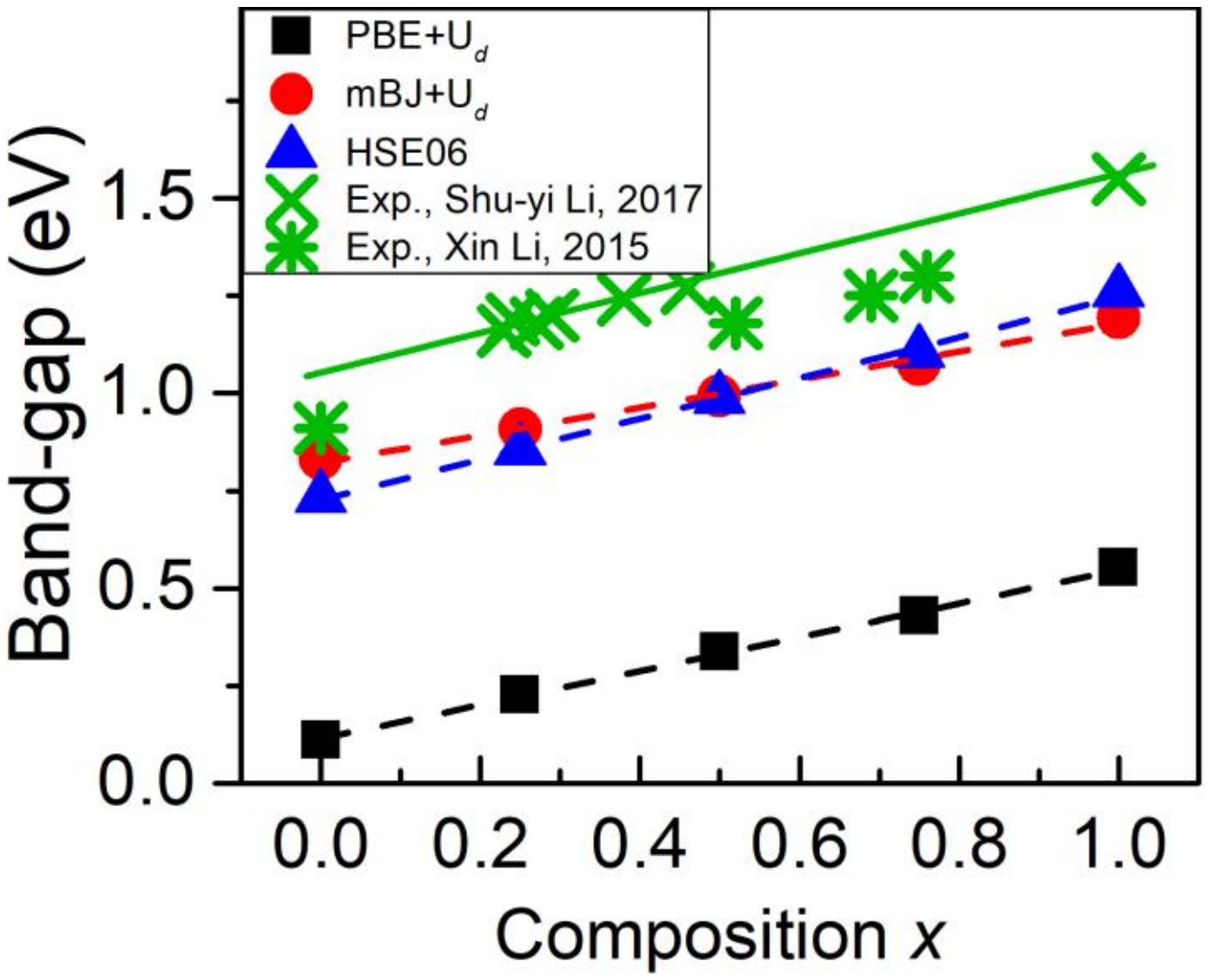

FIG.3. (Color online) Experimental [21,48] and calculated band-gap energies of $Cu_2ZnSn(S_xSe_{1-x})_4$ solid solutions as a function of composition *x*. The full curve shows the corresponding fitting curves of PBE+$U_d$, mBJ+$U_d$ and HSE06 hybrid functionals, as well as the experimental curve of Shu-Li [21].

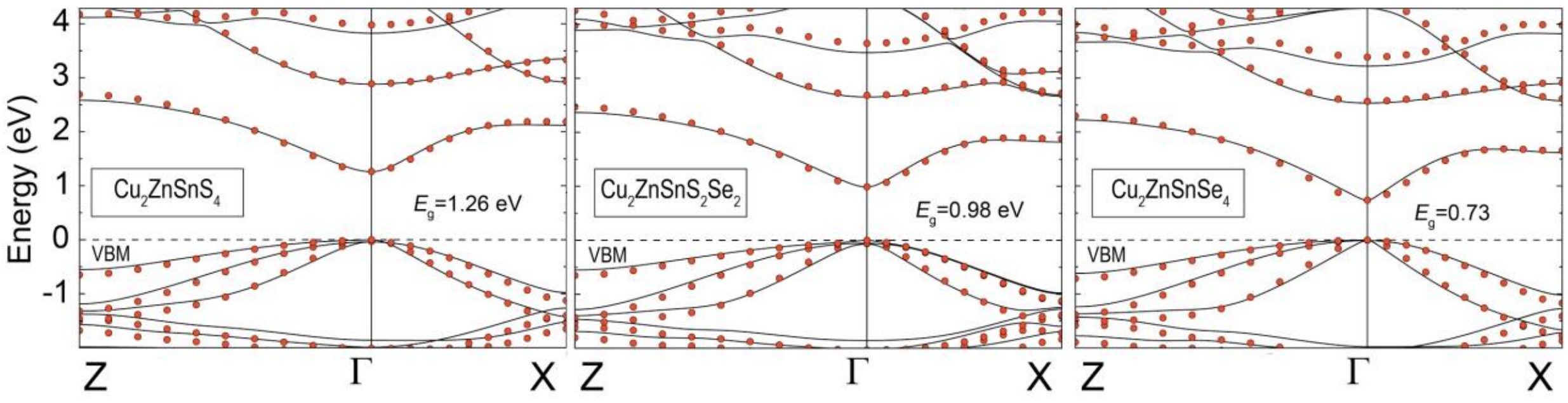

FIG.4. (Color online) The electronic band structures of the $Cu_2ZnSn(S_xSe_{1-x})_4$, with $x$ = 0, 0.5 and 1, of PBE+$U_d$ (solid curves) and HSE06 (dots). The valence band maximum (VBM; dashed lines) is set to zero, while the conduction band minimum of PBE+$U_d$ is shifted to match that of HSE06.

Figure 6 compares the experimental and computed dielectric functions of $Cu_2ZnSnS_4$, here including constant shifts $\Delta_g$ widening the band-gaps to the experimental gaps. The filled areas indicate the experimental results [21]. Both the PBE+$U_d$+$\Delta_g$ results and the mBJ+$U_d$+$\Delta_g$ dielectric functions show the same general features as the experimental data. However, PBE+$U_d$+$\Delta_g$ overestimates the magnitude of both the real and imaginary dielectric function. For example, the magnitude of the mBJ+$U_d$+$\Delta_g$ dielectric function of $Cu_2ZnSnS_4$, (Fig. 6) match better with experiment than PBE+$U_d$+$\Delta_g$, but it still underestimates the peak position of $\varepsilon_2$ by 0.6 eV and likewise for the peaks (or kink) of $\varepsilon_1$ at 3.0 eV. The coarsely (8 × 8 × 8) sampled HSE06+$\Delta_g$ results are also shown and compared with the dielectric function generated with a 24 × 24 × 24 generated using the $\mathbf{k}\cdot\tilde{\mathbf{p}}$ method. Like in Fig. 2, coarse sampling of the mesh cause $\varepsilon_2$ to be underestimated close to the band edge. Moreover, going from calculations based on a coarse mesh to the interpolated dense mesh shifts the energy difference between the two peaks of $\varepsilon_1$ from 0.69 eV to 1.15 eV, which can be compared to the experimental difference of 1.13 eV. This severe underestimation of the peak-to-peak separation shows that a dense **k**-mesh is needed in quantitative comparisons of experimental and calculated dielectric spectrums. Overall, we find that the interpolated HSE06+$\Delta_g$ results (solid curve) more accurately reproduces the experimental peak position $\varepsilon_2$, peaks or kinks of $\varepsilon_1$ and overall shape than what PBE+$U_d$+$\Delta_g$ and mBJ+$U_d$+$\Delta_g$ does.

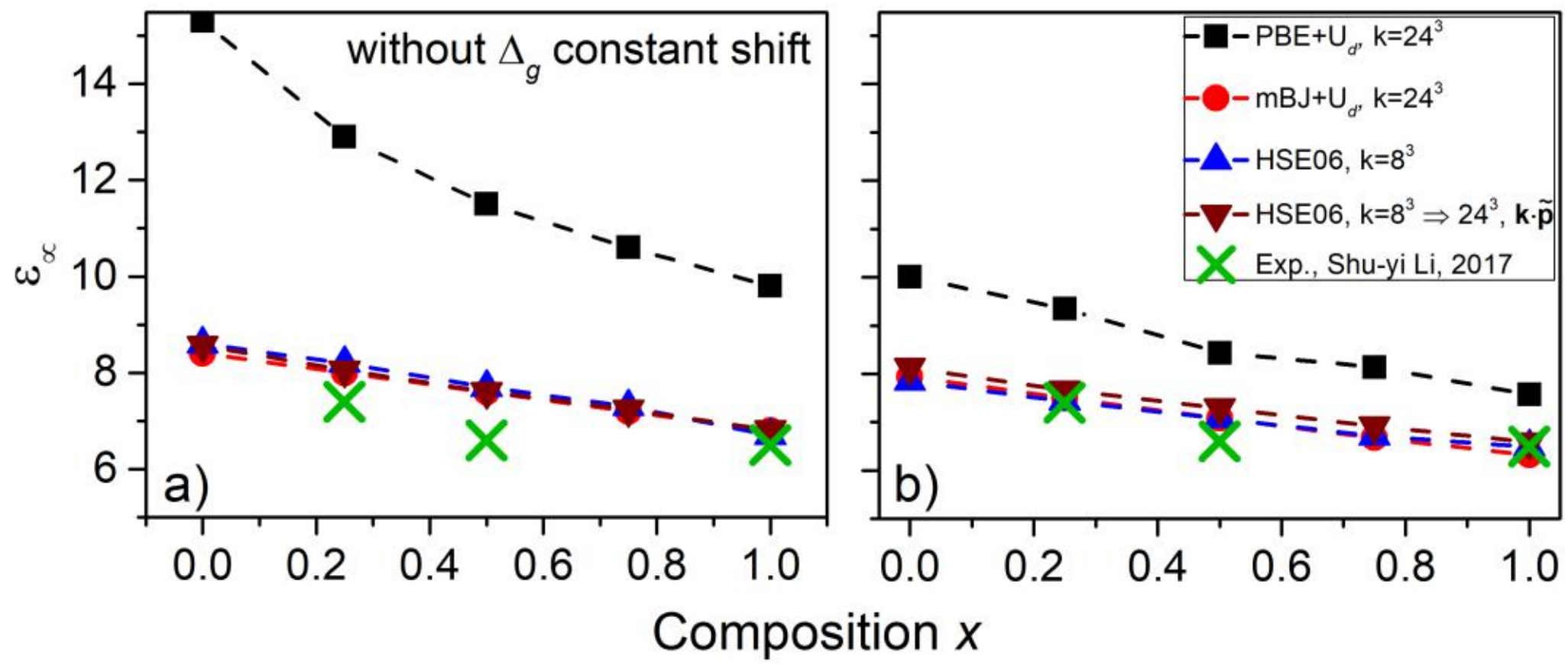

FIG.5. (Color online) Experimental and theoretical high-frequency dielectric constants as a function of composition *x*. The calculations performed with HSE06, PBE+$U_d$ and mBJ+$U_d$ without (a) and with (b) $\Delta_g$ constant shifts.

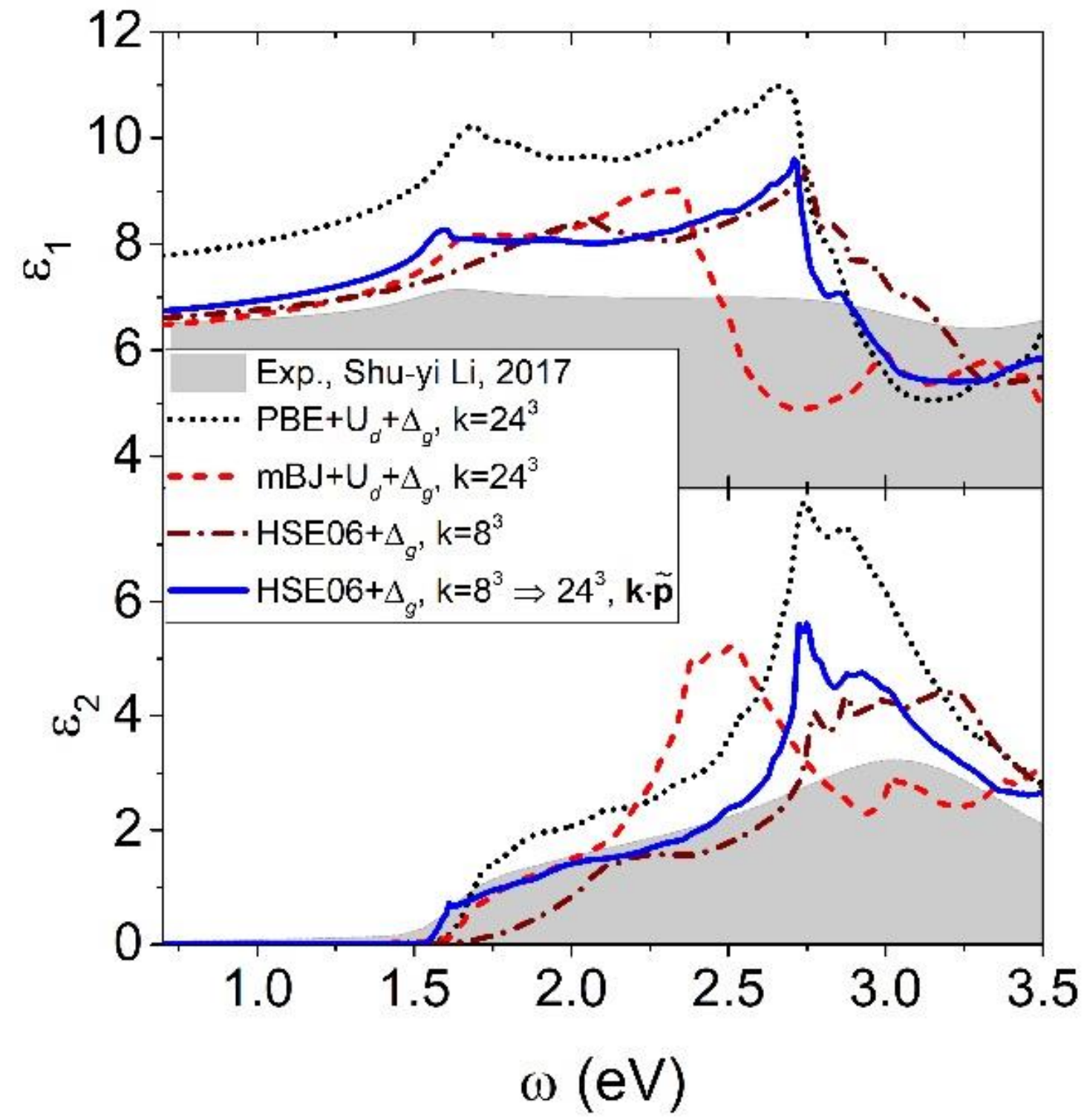

FIG.6. (Color online).The dielectric functions $\varepsilon_1$ and $\varepsilon_2$ of the $Cu_2ZnSnS_4$, as determined experimentally by ellipsometry [21], and first-principles calculations used PBE+$U_d$, mBJ+$U_d$ and HSE06 with the band-gap shifted using constant shifts to the experimental ($\Delta_g$).

Figure 7 provides a quantitative comparison of experimental and calculated positions of the peak positions in the imaginary part of the dielectric function. The first experimental peak of imaginary part of the dielectric function $\varepsilon_2$ is appearing at 3.09 eV [21] (Fig 7(a)) The peak positions calculated by PBE+$U_d$+$\Delta_g$, HSE06+$\Delta_g$, and HSE06+$\Delta_g$ with $\mathbf{k} \cdot \tilde{\mathbf{p}}$ interpolation are all within 0.48 of of experimental value. At the same time, the peak positions of $\varepsilon_2$ curves, calculated with mBJ+$U_d$+$\Delta_g$ (Fig 7 (a)) underestimate experimental value by about 0.8 eV.

An accurate description of peak positions of $\varepsilon_2$ is important, because it relates to the generation profile of the device via absorption coefficient and is it therefore important in assessing the device performance. mBJ+$U_d$+$\Delta_g$ also underestimates the second kink position in the imaginary dielectric function $\varepsilon_2$ (Fig. 7(b)) also but the positions of first kinks are consistent with experimental observation. PBE+$U_d$+$\Delta_g$, and HSE06+$\Delta_g$ with $\mathbf{k} \cdot \tilde{\mathbf{p}}$ interpolation also describe peak and kink positions well. While PBE+$U_d$+$\Delta_g$ is describing critical points and kinks with reasonable accuracy (Fig.7), the dielectric constants $\varepsilon_0$ and $\varepsilon_\infty$ as well as amplitude of the dielectric functions are overestimated (Fig. 5). HSE06+$\Delta_g$ with coarse grid sampling, on the other hand, predicts the $\varepsilon_0$ and $\varepsilon_\infty$ accurately, but fails to calculate the position of the first kink in the imaginary part of the dielectric function. Our results show that HSE06 can with a converged grid sampling describe the various features of the dielectric function better than what semi-local exchenge-corelation functionals can, in line with what one would expect for the general higher accuracy of hybrid functional calculations.

Having established that HSE06 calculations based on a dense mesh are well suited to reproduce the dielectric function of $Cu_2ZnSnS_4$, we compare in Fig. 8 the experimental and theoretical results for three different *x*. While our supercell approach cannot provide compositions identical with the experiment, the concentrations of experiment and theory differ by no more than 0.04 in this comparison.

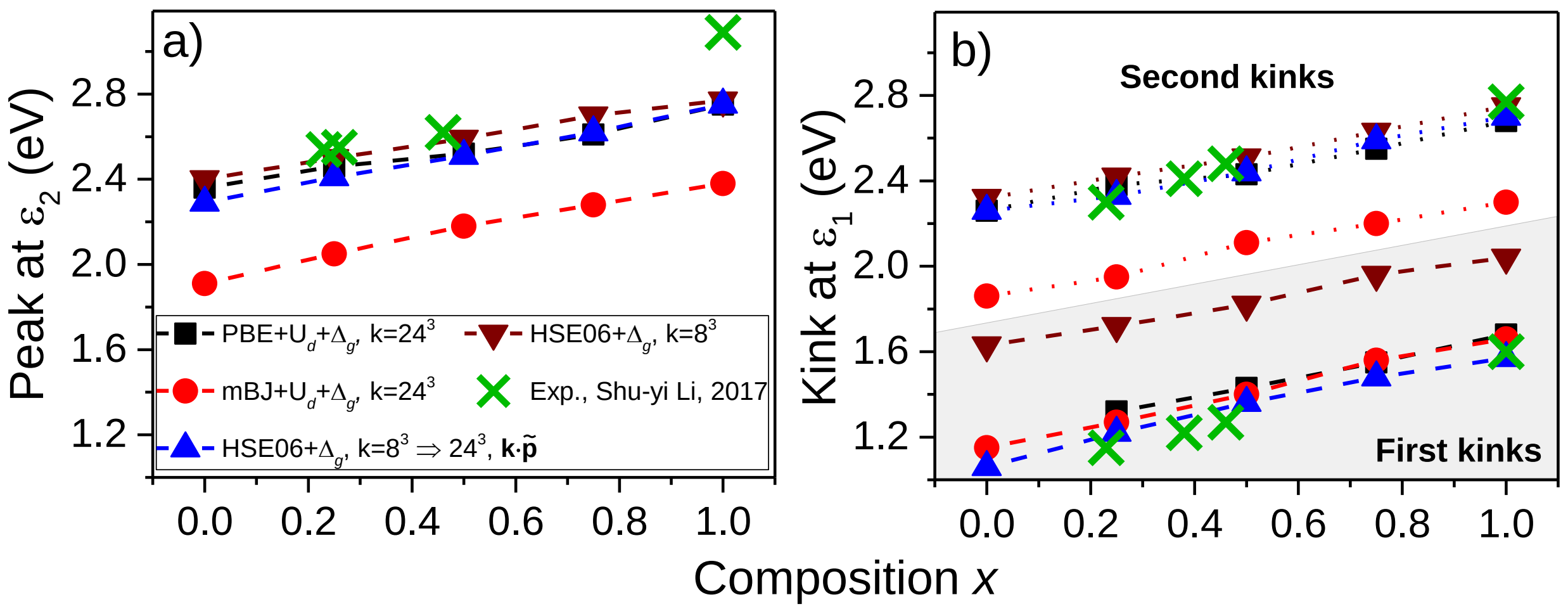

FIG.7. (Color online) Positions of peaks for calculated and experimental dielectric curves $\varepsilon_2$ (a) and kinks for calculated and experimental dielectric curves $\varepsilon_1$ (b) as a function of composition *x*. The calculations performed with PBE+$U_d$+$\Delta_g$, mBJ+$U_d$+$\Delta_g$, HSE06+$\Delta_g$, and HSE06+$\Delta_g$+ $\mathbf{k}\cdot\tilde{\mathbf{p}}$ interpolation.

As shown in Fig 6-8, HSE06 calculations generated with a dense mesh describe the dielectric constants, peak position and critical points of the experimental curves. Figure 8 shows, that the overall shape is quite well described as well. The fact that the experimental curves are far smoother than theory can be due to a number of reasons, including nonhomogeneity of samples and temperature effects. The lower magnitude of the dielectric function of $x$ = 0.46 can be attributed to differences in grain size, and thereby possibly less dense material, for the more sulfur-rich material [21]. Note that even if the magnitudes of $\varepsilon_1$ and $\varepsilon_2$ for $x$ = 0.46 are smaller in experiment compared to theory, the peak/kink positions follow the same trend as the theoretical data.

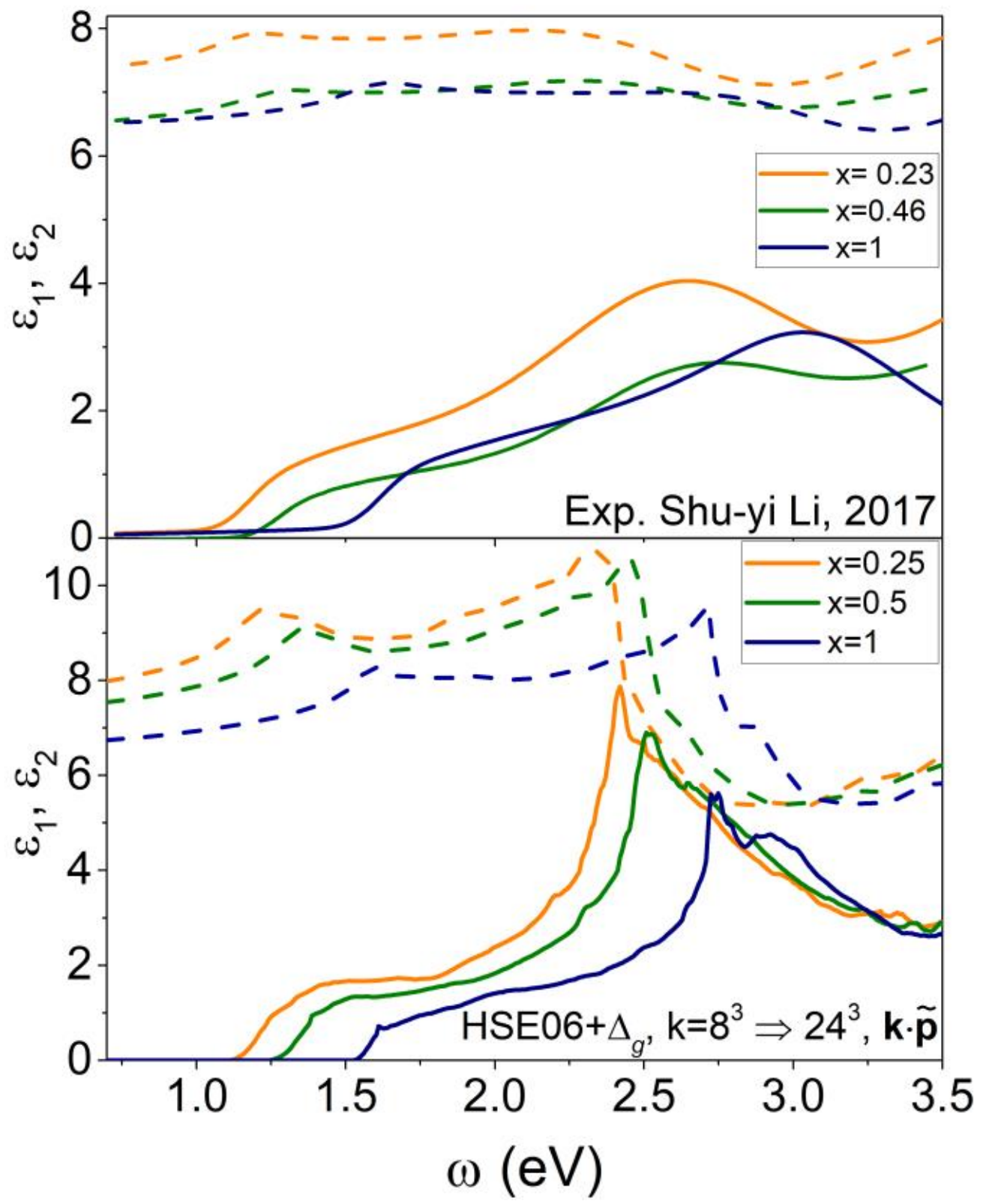

FIG.8. (Color online). The real part $\varepsilon_1$ (dashed curve) and the imaginary part $\varepsilon_2$ (solid curve) of the dielectric functions $\varepsilon = \varepsilon_1 + i\varepsilon_2$ of the $Cu_2ZnSn(S_xSe_{1-x})_4$ (for $x$ = 0.23, 0.46 and 1) determined by experimentally [21] (upper panel) and calculated based on first-principle calculations of $Cu_2ZnSn(S_xSe_{1-x})_4$ (for $x$ = 0.25, 0.5 and 1) with HSE06+$\Delta_g$ (lower panel), including band structure interpolation.

## CONCLUSIONS

A first-principles study of structural, electronic and optical properties of kesterite $Cu_2ZnSn(S_xSe_{1-x})_4$ for different S concentrations has been conducted at different levels of theory. The lattice constants $a$ and $c$, mixing enthalpy, dielectric functions, band-gaps $E_g$ are found to increase monotonically with increasing sulfur anion content $x$. The dielectric constants $\varepsilon_0$ and $\varepsilon_\infty$ decrease with increasing $x$. The calculated electronic band structures of different $Cu_2ZnSn(S_xSe_{1-x})_4$ compositions are qualitatively similar. $\varepsilon_0$ and $\varepsilon_\infty$ of mBJ+$U_d$ and HSE06 agree well with experiment, whereas PBE+$U_d$ overestimates them. Comparing dielectric functions, we find that only HSE06 results based on a densely interpolated to a dense **k**-point sampling of the Brillouin zone is able to both provide accurate magnitudes of the dielectric function and well reproduce peak and kink position of both the real and imaginary part of the experimental dielectric function. The interpolation was achieved by using the recently developed $\mathbf{k}\cdot\tilde{\mathbf{p}}$ method. Our study illustrates the utility of the $\mathbf{k}\cdot\tilde{\mathbf{p}}$ method for accurately computing dielectric functions, when employing methods such as hybrid functionals where computational costs limits one to solving the Kohn-Sham equations with a coarse Brillouin zone sampling. The method therefore enables theory to provide a better and more quantitatively accurate analysis of the experimental dielectric function, which can yield new insights relevant for optimizing the performance of kesterite and related solar cell materials

Finally, we note that there are a number of materials which are characterized by multiple valleys and band contributing to the dielectric functions. For such materials, dense sampling with accurate methods such as hybrid functionals is needed in analyzing optical spectra [51]. The successful application of the $\mathbf{k}\cdot\tilde{\mathbf{p}}$ interpolation methods offer exciting prospects for shedding new insight into the optical properties of such materials.

## ACKNOWLEDGMENTS

This work is supported by the Research Council of Norway (projects 243642 and 228854) and Swedish Foundation for Strategic Research. We acknowledge HPC resources at NSC and USIT through SNIC and NOTUR. We thank Prof. C. Platzer-Björkman, Dr. O. Malyi, and Dr. S. Li for useful discussions.

# Supplementary materials

**Optical properties of $Cu_2ZnSn(S_xSe_{1-x})_4$ by first-principles calculations**

Sergiy Zamulko[1], Kristian Berland[1], and Clas Persson[1,2]
[1]Centre for Materials Science and Nanotechnology, University of Oslo, P. O. Box 1048 Blindern, NO-0316 Oslo, Norway
[2]Department of Materials Science and Engineering, KTH Royal Institute of Technology, Brinellvägen 23, SE-100 44 Stockholm, Sweden

Email address of the corresponding author: sergii.zamulko@smn.uio.no

**Table S1.** Calculated band-gap $E_g$, static $\varepsilon_0$ and high-frequency $\varepsilon_\infty$ dielectric constants of $Cu_2ZnSn(S_xSe_{x-1})_4$ for the alloy composition $x$ = 0, 0.25, 0.5, 0.75 and 1, obtained from modeling 8 atom primitive cells. The $\Delta_g$ corrected $\varepsilon_0$ and $\varepsilon_\infty$ are presented first (before brackets) with gaps fixed to the experimental linear fit of Li *et al.* [21], which are fitted to the experimental gaps indicated with boldface.

| Compound | Method | $E_g$ [eV] | $\varepsilon_\infty$ | $\varepsilon_0$ |
|---|---|---|---|---|
| $Cu_2ZnSnS_4$ | PBE+$U_d$ | 0.55 | 7.6 (9.8) | 10.0 (12.2) |
| | mBJ+$U_d$ | 1.19 | 6.3 (6.8) | 8.8 (9.3) |
| | HSE06 | 1.26 | 6.5 (6.7) | 9.0 (9.3) |
| | experiment | 1.45 [27], **1.55** [21] | **6.5** [21], 7.18 [52], 7.31 [53] | |
| | other calc | 1.50 [17], 1.47 [54] | 6.77 [8] | 4.9 [17] |
| $Cu_2ZnSnS_3Se_1$ | PBE+$U_d$ | 0.43 | 8.1 (10.6) | 10.5 (12.9) |
| | mBJ+$U_d$ | 1.10 | 6.6 (7.2) | 9.2 (9.6) |
| | HSE06 | 1.07 | 6.7 (7.3) | 9.3 (9.6) |
| | other calc | 1.35 [17], 1.30 [54] | | 5.1 [17] |
| $Cu_2ZnSnS_2Se_2$ | PBE+$U_d$ | 0.34 | 8.7 (11.5) | 10.8 (13.7) |
| | mBJ+$U_d$ | 0.99 | 7.1 (7.6) | 9.1 (9.8) |
| | HSE06 | 0.99 | 7.1 (7.7) | 9.2 (9.8) |
| | experiment | **~1.27** [21] | **6.6** [21] | |
| | other calc | 1.23 [17], 1.17 [54] | | 5.3 [17] |
| $Cu_2ZnSnS_1Se_3$ | PBE+$U_d$ | 0.22 | 9.3 (12.9) | 11.4 (14.9) |
| | mBJ+$U_d$ | 0.85 | 7.5 (8.0) | 9.5 (10.0) |
| | HSE06 | 0.91 | 7.4 (8.2) | 9.4 (10.1) |
| | experiment | **~1.19** [21] | **7.4** [21] | |
| | other calc | 1.06 [17], 1.07 [54] | | 5.5 [17] |
| $Cu_2ZnSnSe_4$ | PBE+$U_d$ | 0.11 | 10.0 (15.3) | 12.5 (17.9) |
| | mBJ+$U_d$ | 0.83 | 7.9 (8.4) | 10.7 (10.9) |
| | HSE06 | 0.74 | 7.8 (8.6) | 10.5 (11.2) |
| | experiment | 1.00 [27] | 8.87 [55] | |
| | other calc | 1.00 [17], 0.9 [54] | | 5.6 [17] |

**Table S2.** Calculated constant shift $\Delta_g$ for the $Cu_2ZnSn(S_xSe_{x-1})_4$ alloy composition $x$ = 0, 0.25, 0.5, 0.75 and 1 and functionals (HSE06, PBE+$U_d$ and mBJ+$U_d$).

| Concentration $x$ of S | $\Delta_g$, [eV] | | |
|---|---|---|---|
| | mBJ+$U_d$ | PBE+$U_d$ | HSE06 |
| 0 | 0.936 | 0.217 | 0.315 |
| 0.25 | 0.951 | 0.270 | 0.324 |
| 0.5 | 0.970 | 0.315 | 0.322 |
| 0.75 | 1.008 | 0.364 | 0.334 |
| 1 | 1.015 | 0.374 | 0.308 |

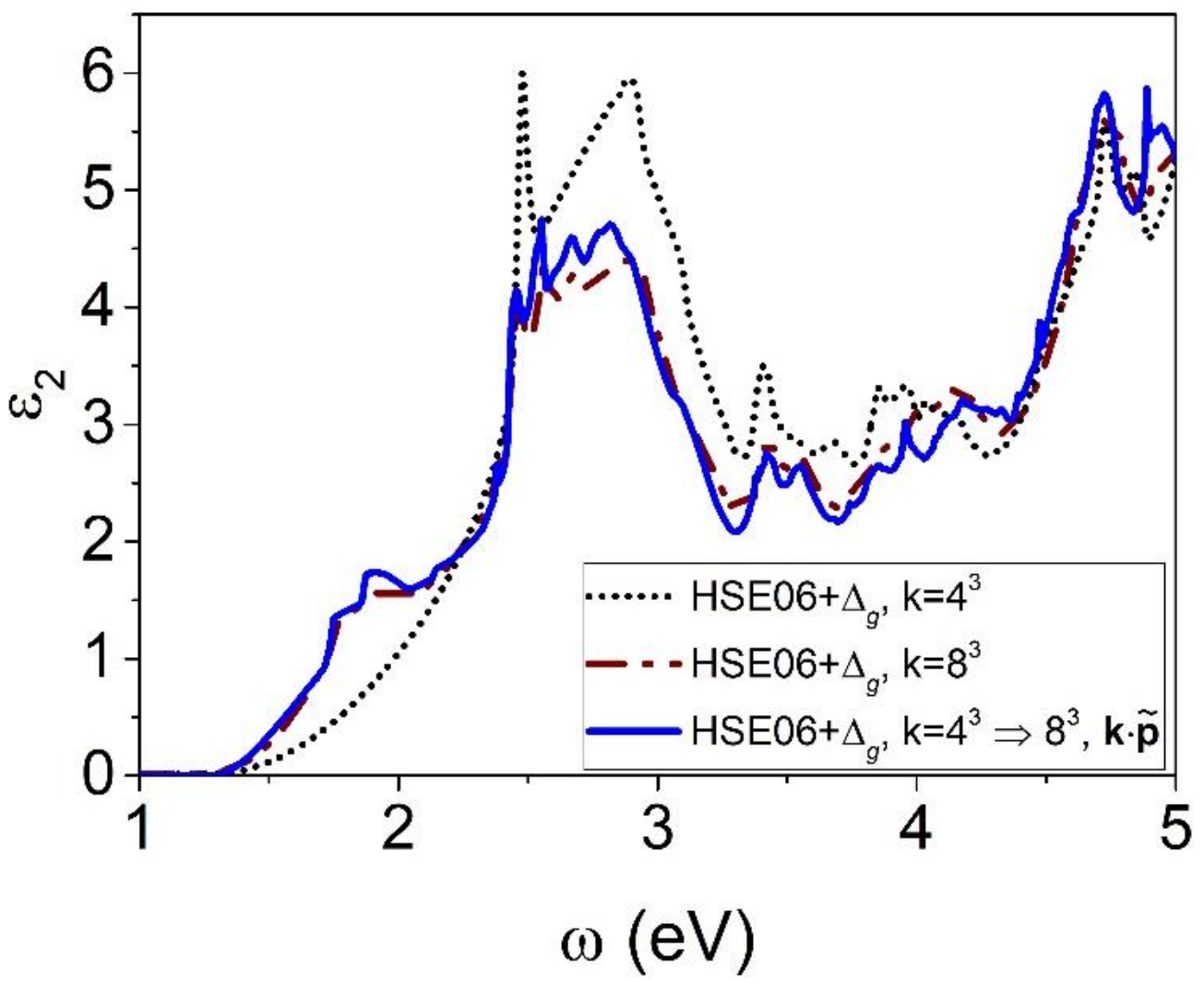

FIG.S1. (Color online) The dielectric function of $Cu_2ZnSnS_4$ calculated with HSE06 with an $4 \times 4 \times 4$ ($\mathbf{k} = 4^3$) and $8 \times 8 \times 8$ ($\mathbf{k} = 8^3$) Γ-centered **k**-point sampling. The $\mathbf{k} \cdot \tilde{\mathbf{p}}$ method is used to extend **k**-grid form $\mathbf{k} = 4^3$ to $\mathbf{k} = 8^3$.

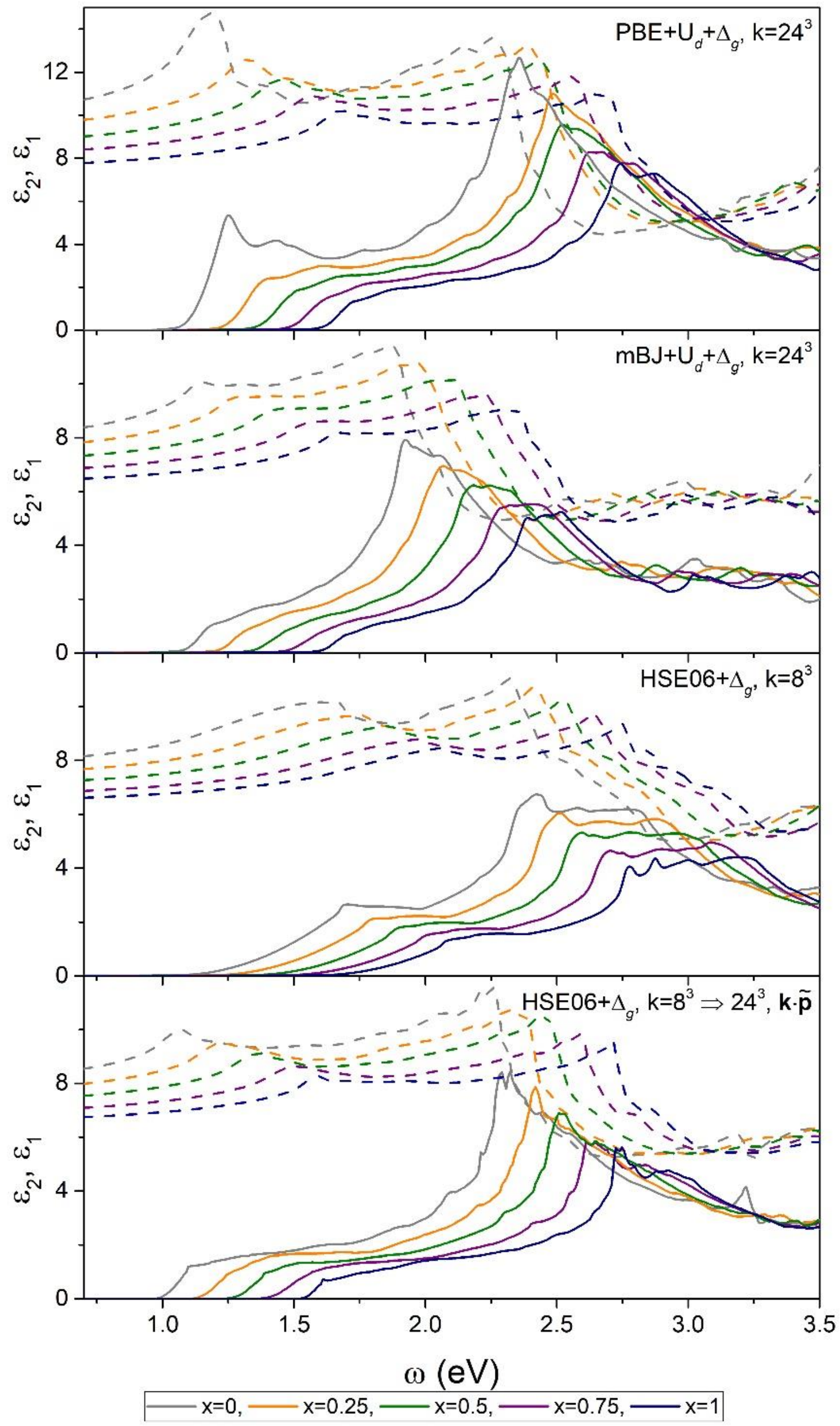

FIG.S2. (Color online). The dielectric functions ($\varepsilon_1$ (dashed curve) and $\varepsilon_2$ (solid curve)) of the $Cu_2ZnSn(S_xSe_{1-x})_4$ determined by experimental ellipsometry ([21]), and extended by $\mathbf{k}\cdot\tilde{\mathbf{p}}$ method first-principles calculations with HSE06 functional Here, for comparison, the constant shift ($\Delta_g$) have been used.